\documentclass[a4paper]{ESASPCS13Style}
\usepackage{epsfig}

\begin{document}

\title{Current Status of the SuperWASP Project}
\author{D.J. Christian\inst{1}, D.L. Pollacco\inst{1},
W.I. Clarkson\inst{2},
A. Collier Cameron\inst{3},
N. Evans\inst{4},
A. Fitzsimmons\inst{1},
C.A. Haswell\inst{2},
C. Hellier\inst{4},
S.T. Hodgkin\inst{5},
K. Horne\inst{3},
S.R. Kane\inst{3},
F.P. Keenan\inst{1},
T.A. Lister\inst{3},
A.J. Norton\inst{2},
R. Ryans\inst{1},
I. Skillen\inst{6},
R.A. Street\inst{1},
R.G. West\inst{7},
P.J. Wheatley\inst{7}
}

\institute{APS Division, Department of Pure \& Applied Physics, Queen's University Belfast, Belfast, BT7 1NN, UK
\and
Department of Physics \& Astronomy, The Open University, Milton Keynes, MK7 6AA, UK
\and
School of Physics \& Astronomy, University of St. Andrews, North Haugh, St. Andrews, Fife, KY16 9SS, UK
\and
Astrophysics Group, School of Chemistry \& Physics, Keele University, Staffordshire, ST5 5BG, UK 
\and
Institute of Astronomy, University of Cambridge, Madingley Road, Cambridge, CB3 0HA, UK
\and
Isaac Newton Group of Telescopes, Apartado de correos 321, E-38700 Santa Cruz de la Palma, Tenerife, Spain 
\and
Department of Physics \& Astronomy, University of Leicester, Leicester, LE1 7RH, UK 
}

\maketitle 

\begin{abstract}

 We present the current status of the SuperWASP project, a Wide Angle Search
for Planets.  SuperWASP consists of up to 8 individual cameras using
ultra-wide field lenses backed by high-quality passively cooled CCDs. 
Each camera
covers 7.8 x 7.8 sq degrees of sky, for nearly 500 sq degrees of sky coverage.
SuperWASP I, located in LaPalma, is currently operational with
5 cameras and is conducting a photometric survey of a large numbers
of stars in the magnitude range $\sim$7 to 15.  The collaboration has 
developed a custom-built reduction pipeline
and aims to achieve better than 1 percent photometric precision.
The pipeline will also produce well sampled light curves for all
the stars in each field which will be used to detect:
planetary transits, optical transients, and track Near-Earth Objects.
Status of current observations, and expected rates of extrasolar planetary 
detections
will be presented. The consortium members, institutions, and further details
can be found on the web site at: http://www.superwasp.org.

\end{abstract}

\section{Introduction}
  
The search for extrasolar planets (ESP) has been dominated by radial velocity 
surveys (\cite{MQ95}; \cite{MB98}; \cite{Ud00}). 
These surveys have found Jupiter-sized planets orbiting 6\% of nearby Sun-like 
stars.  About 1/3 of these exoplanets have been found in very close orbits 
(a $\approx$ 0.05 AU) with periods $<$ 4 days. A transit of such a hot-Jupiter 
will dim the host star by $\sim$0.01 mags. This was detected for G-star (V=7.7)
HD 209458 (\cite{ch00}). Recently, 3 transiting candidates from the
Optical Gravitational Lensing Experiment have been confirmed as ESP 
(\cite{kon03}; \cite{kon04}; \cite{bo04}), as well as, the first
ESP from the Trans-Atlantic Exoplanet Survey network (\cite{al04}).

Future transit detections will provide the planet's radius, mass, and density. 
The expected number of such transits, ranges from 1 (\cite{B03}) to a few per 
20,000 stars surveyed. Follow-up observations will be needed to separate true 
transit events from false detections, such as grazing eclipses in a system of 
two main-sequence stars. There are currently nearly two dozen photometric 
surveys underway (\cite{H03}), and here we present the current status of the 
Wide Angle Search for Planets (called SuperWASP).  SuperWASP covers a wider 
area of sky and more stars than any other current survey.

\subsection{Science Objectives}
SuperWASP was designed to cover a large area of sky and achieve 
photometric accuracy of a few millimags. These specifications
make SuperWASP ideal for monitoring several 10's of thousands
of stars to search for transiting extrasolar planets,
discovery and follow-up of optical transients (including
optical flashed from $\gamma$-ray bursts), and the discovery
and tracking of near-Earth objects and asteroids.
Naturally SuperWASP is also well suited for a variety
of wide field imaging and temporal studies, such as 
discovery and monitoring of variable stars, stellar flares 
and other periodic and
aperiodic celestial phenomenon. First results from the
study of variable stars in the Pleadies have
been presented by \cite*{L04}. 

\section{Instrumentation}

 The SuperWASP~I telescope in LaPalma was designed based on the
success of the prototype WASP0 instrument (\cite{Kane04}).
The SuperWASP telescope in LaPalma is contained it is own
commercially available enclosure with a hydraulically operated 
roll-away roof, and its own GPS and weather station. Commercially available 
components were used where available to keep costs down
and decrease construction time.  The telescope mount is
a rapid slewing fork mount ($\approx$10$^{o}$/sec).
To meet the science requirements of a covering a large area
of sky with high photometric precisions the combination of 
Canon 200-mm f/1.8 lenses and Andor e2v 2kx2k back illuminated CCDs
was chosen. The CCDs are passively cooled with a Peltier cooler
and have a very short 4-sec readout time. This combination of
lens and camera gives a field of view of 7.8$^o$ x 7.8$^o$ 
($\sim$61 sq degrees).
This set-up covers from 7--15 magnitude for typical 30 seconds exposure
and can achieve $<$1\% photometric precision for magnitudes up to 12.
The complete instrument will have 8 cameras and cover 490 sq degrees!
The current observing configuration for 2004 is with 5 cameras (Figure~1.). 

\begin{figure}[ht]
  \begin{center}
    \epsfig{file=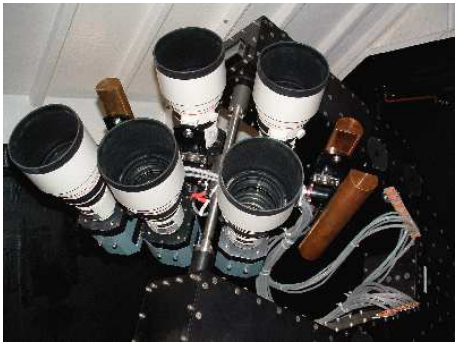, width=6cm}
  \end{center}
\caption{The 5 SuperWASP Cameras in LaPalma.
\label{fig1}}
\end{figure}

\subsection{Observational Strategy}
 The SuperWASP instruments were commissioned and 'on-sky'
in November 2003. Observations were resumed after the 
winter break and the telescope was inaugurated on 16 April 2004.
Observers were present for the 2004 observing season to monitor
the systems and make improvements as the telescope moved from
automated to robotic operations. 
The initial observing strategy was tailored toward searching
for exoplanet transits.  This strategy was to 
observe fields with a large number of stars, but to avoid
the Galactic plane where over crowded fields would make reductions
impossible. Based on the Besancon galactic model a declination 
of +28 was chosen
with the telescope stepping through RA centered on the current
LST but within the $\pm$4.5 hr hour angle limit of the mount. A maximum
of 8 fields are observed with a duration of $\approx$1 minute per
field. This 1 minute includes the 30 second exposure, 4 seconds read-out and
the time for the telescope to slew and settle at the field. 
Such observations provide well sampled light curves with a maximum
of 8 minutes per measurement. 
This observing strategy provides over 6 hours of coverage for 
the 6 fields centered on LST at midnight, and over 4 hrs of coverage
on 10 fields per night. 

 A typical field contains $\sim$25,000 stars per camera at
magnitudes brighter than 13. Considering that only
14--19\% of these stars are late-type (F--M) stars,
we therefore expect $\sim$4000 F-M dwarf stars per field per camera.
If 1\% of these have hot-Jupiter companions (\cite{LG03}) 
and we expect 10\% of these 
to show transits (\cite{H03}),  we therefore expect to detect 
$\approx$4 transits per camera per field for $\approx$40 clear nights of
monitoring (see Table~1). This would give 120 to 
200 transit candidates for all 5 cameras from monitoring 6 to 10 fields, 
respectively.

There are many different types of stellar systems that can 
show behavior mimicking a planets transit (see Brown 2003 
for example). 
Therefore, transit candidates will be followed-up with 
additional optical photometry and low resolution spectroscopy to confirm
that a planet is the true cause of the transit.
Deeper imaging on 1 -- 2 meter class telescope will be used to 
confirm that the transit is not caused by a foreground or background
grazing eclipsing system and to also obtain photometric colors for
determining spectral type.
The transit depth is inversely proportional to the
radius of the primary star squared (\cite{H03}). Therefore,
the transit depth will be deepest for the later
type stars. 
If the transit candidate passes these initial follow-up observational
tests then higher resolution (few m/s) radial velocity observation will 
be obtained to derive the orbital parameters. 

\begin{figure}[ht]
  \begin{center}
    \epsfig{file=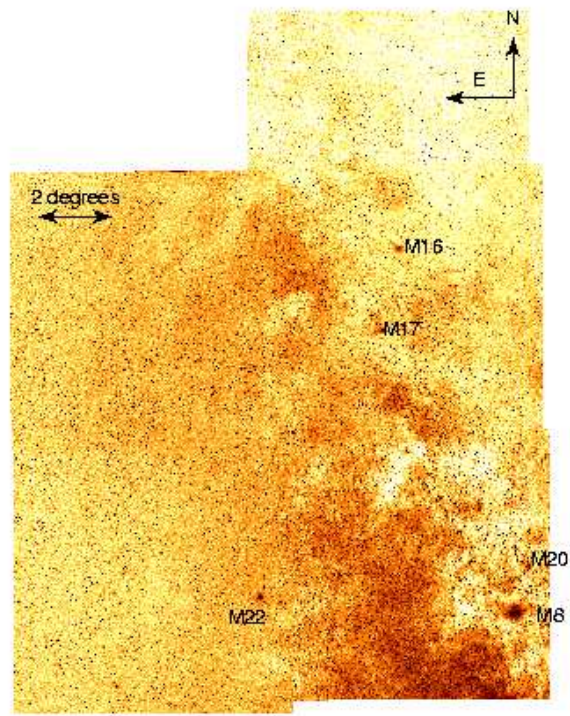, width=8cm}
  \end{center}
\caption{SuperWASP 5 Camera mosaic centered near M25 (RA 18:31, Dec -19:15).
Image dimensions are $\sim$15$^o$ x 20$^o$ and exposure time is 1 second.
Several Messier objects are noted.
\label{fig2}}
\end{figure}

\begin{table}
\begin{center}
\caption{Number of extrasolar planets and expected number of transits}
\label{transits}
    \leavevmode
    \footnotesize
\begin{tabular}{lc}
\hline
\hline
        Number of ESP & Reference      \\\hline
   300 per 25,000$^a$ &    Lineweaver \& Grether (2003)    \\
   100 per 2000       &    current Doppler Surveys     \\
\\
 ESP Transit Predictions  &          \\\hline
   1 per 25,000           & Brown 2003 \\
   9 per 25,000           & Horne (2003) \\ 
   4 per 25,000           &  This work \\ \hline

\end{tabular}
\end{center}
$^{a}$5--9\% of Sun-like stars \\
\end{table}

\section{Data Reduction}
An individual camera image is 8.4 MB and stored in FITS format.
About
600 images are obtained per clear night per camera for
a typical 9 hours of observing (including calibrations).
Thus, for 8 cameras about $\sim$40 GB/per night is collected 
and stored on DLT before being ingested into the custom-built
data reductions pipeline (\cite{S04}). Data are compressed with
the {\it cfitsio} routine {\it imcopy}, which uses the
lossless Rice algorithm and obtains up to a factor of $\approx$2 compression.

\subsection{Pipeline}
The collaboration has built a 
custom data reduction\\
 pipeline with the goal of
obtaining a few millimag photometric precision 
for stars with V $<$ 12. 
The pipeline uses custom written fortran programs combined
with shell scripts and several STARLINK packages.
The pipeline creates master biases, darks, and flat fields
for each night of observations. Each science exposure 
is bias subtracted, dark corrected, flat fielded,
and the astrometric solution is computed using reference stars
from the USNO-B1.0 and Tycho catalogs.   
Aperture photometry is performed with a package custom-built to 
deal with the ultra-wide fields. Bad pixel masks are applied to
each frame and a blending index is assigned for every object detected. 
Fluxes are computed in
3 different apertures and outputted, along with other source attributes
to a FITS binary table. Trends (such as airmass) are removed from the
data and the binary table is ingested by the archive.

\subsection{Archive}
This public archive is hosted at Leicester within 
LEDAS (Leicester Database and Archive Service) and can be 
accessed at http://www.ledas.ac.uk/ 
A list of all cataloged targets is complied from the
ingested FITS tables and compared with the WASP catalog.
New objects are assigned IAU-compatible names and all objects
are assigned to a 5$^o$x5$^o$ sky tiles based on their coordinates.
Photometric data are split from the input files and stored
as intermediate per-sky-tiles. Photometric points within each 
sky-tile are re-ordered to insure that they are in consecutive rows
for a given star. These files are registered within the archive
DBMS along with their names, location, and various meta-data. 
The archive can then be interrogated and the photometric data
extracted as a single, coherent per-object light curve for
longer term temporal analysis. The object catalog is expected
to contain $>$ 3x10$^6$ objects per year.  

\section{Future Goals}

 SuperWASP~I in LaPalma will be fully robotic for the 2005
observing season. All-sky survey and transient follow-up
observations will be added to the planet transit search
program. Improvements to the transient follow-up observations will 
include real-time quick reduction of images with the coordinates of any new 
transient sent to the Liverpool telescope for real-time follow-up.
A sister telescope to SuperWASP~I in LaPalma is under construction 
in South Africa and first light is expected in mid-2005.

\begin{acknowledgements}
SuperWASP is a consortium of: Queen's University Belfast, University of Cambridge, 
Isaac Newton Group of Telescopes (La Palma), University of Keele, University of Leicester,
Open University, and the University of St Andrews.
The SuperWASP instrument has been funded by major financial contributions 
from Queen's University Belfast, the Particle Physics and Astronomy Research Council 
and the Open University.
SuperWASP I is located in the Spanish Roque de Los Muchachos Observatory on La Palma, 
Canary Islands which is operated by the Instituto de Astrofísica de Canarias (IAC).

\end{acknowledgements}

\end{document}